\documentclass[%
 reprint,
 amsmath,amssymb,
 aps,
]{revtex4-1}

\usepackage{graphicx}
\usepackage{dcolumn}
\usepackage{bm}
\usepackage{color}
\usepackage[english]{babel}

\begin{document}

\preprint{APS/123-QED}

\title{\bf{Spectroscopy of Equilibrium and Non-Equilibrium Charge Transfer in Semiconductor Quantum Structures}}

\author{C. R{\"{o}}ssler}
 \email{roessler@phys.ethz.ch}
\author{S. Burkhard}
\author{T. Kr{\"{a}}henmann}
\author{M. R{\"{o}}{\"{o}}sli}
\author{P. M{\"{a}}rki}
\author{J. Basset}
\author{T. Ihn}
\author{K. Ensslin}
\author{C. Reichl}
\author{W. Wegscheider}

\affiliation{Solid State Physics Laboratory, ETH Zurich, 8093
Zurich, Switzerland}%

\date{\today}

\begin{abstract}
\noindent We investigate equilibrium and non-equilibrium charge-transfer processes by performing high-resolution transport spectroscopy. Using electrostatically defined quantum dots for energy-selective emission and detection, we achieved unprecedented spectral resolution and a high degree of tunability of relevant experimental parameters. Most importantly, we observe that the spectral width of elastically transferred electrons can be substantially smaller than the linewidth of a thermally broadened Coulomb peak. This finding indicates that the charge-transfer process is fast compared to the electron--phonon interaction time. By drawing an analogy to double quantum dots, we argue that the spectral width of the elastic resonance is determined by the lifetime broadening $h\it{\Gamma}$ of the emitter and detector states. Good agreement with the model is found also in an experiment in which the charge transfer is in the regime $h\it{\Gamma}\gg k_{\rm{B}}T$. By performing spectroscopy below the Fermi energy, we furthermore observe elastic and inelastic transfer of holes.
\end{abstract}

\pacs{Valid PACS appear here}
\maketitle

\noindent Nanostructured devices harboring two-dimensional electron systems (2DESs) form a well-established platform for studying the interactions of electrons among one another and with their environment. Furthermore, the high degree of control and tunability in these systems enables a wide range of fundamental and technological applications exploiting the quantum nature of electrons, ranging from the exploration of topological phases to various uses in quantum information science. Many of these applications require ballistic, non-adiabatic electronic transport over macroscopic distances, during which the quantum state of the electrons is preserved, without thermalization with the reservoir. Here, we describe and demonstrate a novel scheme for such non-equilibrium charge transfer in semiconductor quantum structures, which provides at the same time a spectroscopic tool for discriminating between ballistically and diffusely transported electrons and holes.

Non-equilibrium charge carriers injected into the reservoir of a semiconductor nanostructure undergo complex relaxation mechanisms. In seminal work, the interaction between `hot' electrons and phonons has been probed via emission and detection across barriers~\cite{sivan_hot_1989,dzurak_two-dimensional_1992}. For excess energies $E_{\rm{EX}}$ approaching the longitudinal-optical (LO) phonon energy $E_{\rm{PHONON}}^{\rm{LO}}\approx 36\,\rm{meV}$ (in GaAs), strong relaxation of hot electrons due to LO-phonon emission has been observed. In subsequent magnetic-focusing experiments with crossing ballistic electron beams~\cite{muller_electron-electron_1995}, electron--electron interactions were shown to become dominant at excess energies comparable to the Fermi energy of the 2DES, $E_{\rm{F}}\sim10\,\rm{meV}$. In this regime, the lifetime of hot electrons is well understood~\cite{giuliani_lifetime_1982}; in particular, the scattering rate has been shown to depend on the square of the energy, $E^2$. Consequently, experiments designed to observe ballistic non-equilibrium electron transfer were typically performed at even smaller excess energies, $E_{\rm{EX}}\sim1\,\rm{m eV}$, or involved magnetic fields to reduce back-scattering. For instance, electron relaxation in edge channels in the quantum Hall regime was probed via a side-coupled quantum dot (QD)~\cite{altimiras_non-equilibrium_2010,le_sueur_energy_2010} and thermometry of neutral modes was performed on the quantum Hall edge~\cite{venkatachalam_local_2012}. In pioneering magnetic-focusing experiments with QDs~\cite{hohls_ballistic_2006}, transport across several micrometers has been achieved; specifically, it was shown that electrons reach the detector QD at energies $E\lesssim E_{\rm{EX}}=300\,\rm{\mu e V}$.

In the experiments reported here, we performed high-resolution spectroscopy between two QDs directly, that is, without magnetic field. Operation at small excess energies of $E_{\rm{EX}}\sim100\,\rm{\mu eV}$ minimizes electron--electron interaction and facilitates the detection of elastically transferred electrons. We find that the spectral width of the elastic-transfer resonance can be significantly below the thermal broadening $k_{\rm{B}}T$, indicating that the charge transfer process is fast compared to the electron--phonon interaction time. Furthermore, we demonstrate that our QD-based spectroscopy scheme is applicable to both electrons (if $E_{\rm{EX}}>E_{\rm{0}}$) and holes (if $E_{\rm{EX}}<E_{\rm{0}}$).

\begin{figure}
\includegraphics[scale=1]{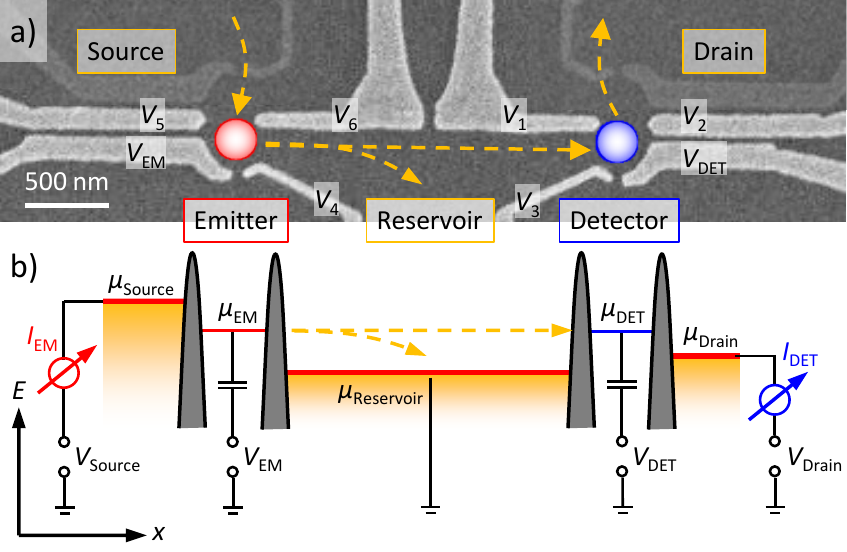}
\caption{\label{fig:sample} a) Scanning electron micrograph of the sample surface. The GaAs crystal appears dark, bright areas mark Schottky gates which deplete the 2DES underneath when a negative voltage is applied to them, whereas gates shown in dark grey are grounded during the experiment. The QDs denoted {\it{emitter}} and {\it{detector}} are separated by $2\,\mu m$ and are defined by choosing appropriate voltages for $V_1$--$V_6$, $V_{\rm{EM}}$ and $V_{\rm{DET}}$. b) Schematic of non-equilibrium electron transfer. Electrons tunnel from the source to the emitter,  where they are injected into the grounded reservoir. The height of the detector ground state determines the energy at which the beam of hot electrons is probed. Current flow into and out of the source ($I_{\rm{EM}}$) and drain ($I_{\rm{DET}}$) is measured independently.}
\end{figure}

\begin{figure}
\includegraphics[scale=1]{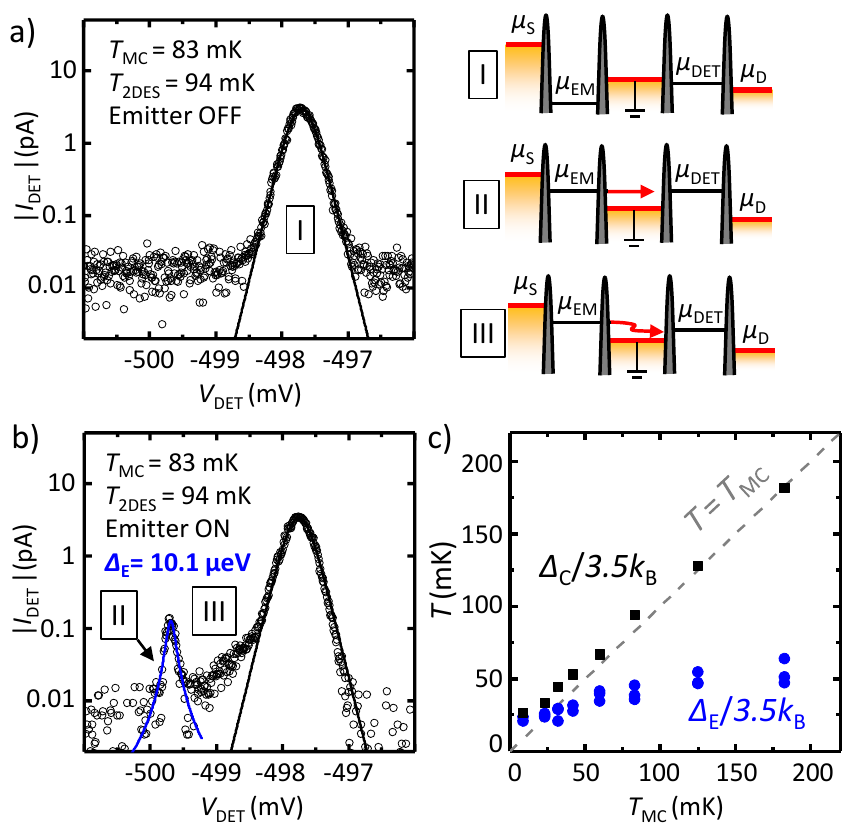}
\caption{\label{fig:sub-thermal} a) Coulomb-blockade oscillation of the detector QD (schematic I), measured at a cryostat temperature of $T_{\rm{MC}}=83\,\rm{mK}$. The black line is a fit assuming thermal broadening with a temperature of $T_{\rm{2DES}}=94\,\rm{mK}$. b) Coulomb-blockade oscillation of the detector QD while the emitter injects electrons at an energy $\mu_{\rm{EM}}= 130\,\rm{\mu eV}$ above the reservoir energy. The blue line is a Lorentzian fit to \texttt{¥}the elastic resonance (schematic II) yielding a FWHM of ${\it{\Delta}}_{\rm{E}}=10.1\,{\rm{\mu eV}}$. Additional current between the Coulomb peak and the elastic resonance originates from inelastically scattered electrons (schematic III). c) Temperature of the Coulomb resonance (squares) and FWHM of the elastic resonance normalized by $3.5k_{\rm{B}}$ (circles) as a function of the cryostat temperature. Whereas $T_{\rm{2DES}}$ changes, as expected, proportionally to $T_{\rm{MC}}$, the elastic linewidth exhibits only a weak temperature dependence and can be much lower than $T_{\rm{MC}}$.}
\end{figure}

\begin{figure}
\includegraphics[scale=1]{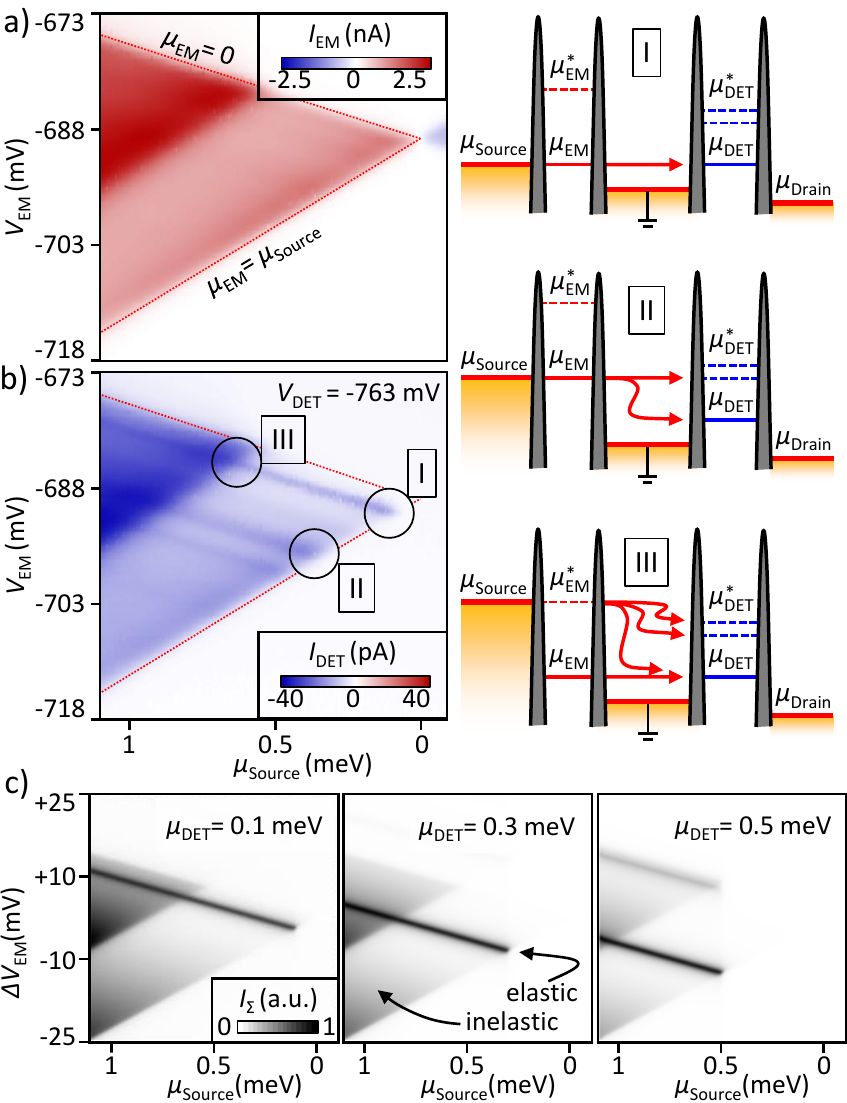}
\caption{\label{fig:finite-bias} a) Emitter current $I_{\rm{EM}}$, plotted in linear color scale as a function of $V_{\rm{EM}}$ and $\mu_{\rm{Source}}$. White areas mark the Coulomb-blockade regime, whereas finite current occurs within the sequential-tunneling regions. A doubling of the emitter current (top left) occurs due to an excited orbital state entering the bias window. b) Simultaneously measured detector current $I_{\rm{DET}}$, while ${\it{\mu}}_{\rm{DET}}$ is slightly above the chemical potential of the reservoir and drain (see sketches on the right-hand side). Finite current is only observed in the regime where the emitter injects non-equilibrium electrons into the reservoir, marked by red dashed lines. The resonance condition between emitter GS and detector GS is met along the streaks labelled I and III; additional streaks can be attributed to detector ESs (labeled II). c) Calculated current through the two-QD system, plotted in linear gray scale as a function of ${\it{\Delta}} V_{\rm{EM}}$ and $\mu_{\rm{Source}}$. Resonance between emitter GS and detector GS occurs along the black streak, current within the filled triangles is caused by inelastic scattering events in the reservoir. The individual panels show setting with different energies of the detector GS (increasing from left to right). At higher detector-GS energies, the onset of transmission shifts to larger $\mu_{\rm{Source}}$.}
\end{figure}

\begin{figure}
\includegraphics[scale=0.9]{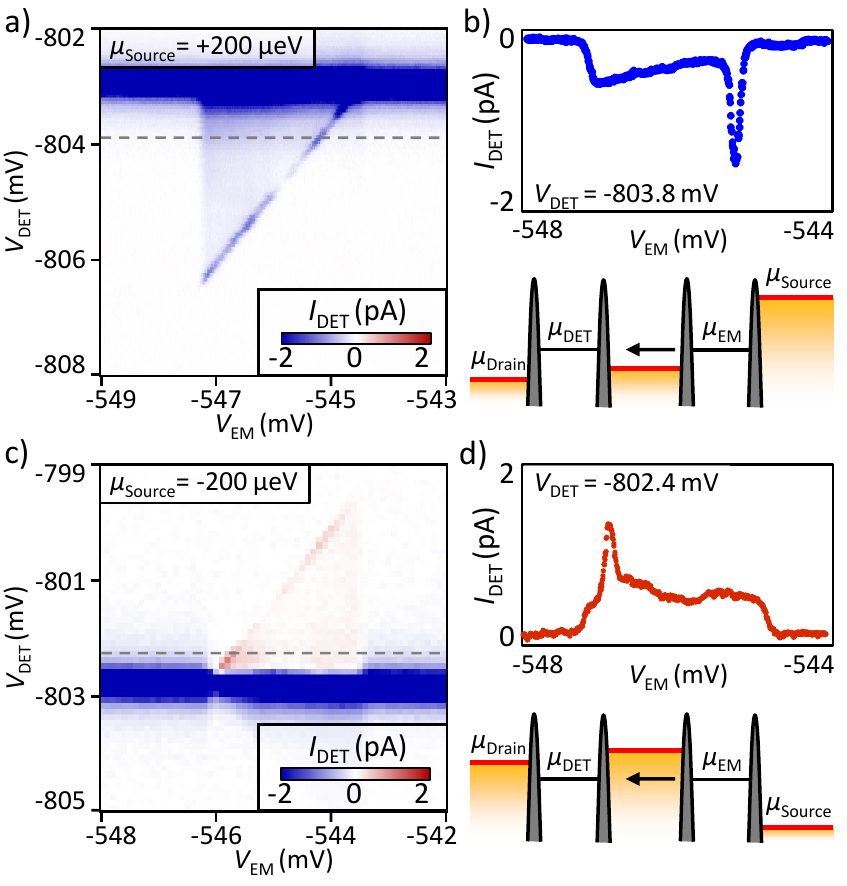}
\caption{\label{fig:hole-transfer} a) Detector current $I_{\rm{DET}}$, plotted in linear color scale as a function of $V_{\rm{EM}}$ and $V_{\rm{DET}}$. Blue denotes negative current. Along the blue diagonal line, electrons tunnel from source to $\mu_{\rm{EM}}$, are elastically transferred to $\mu_{\rm{DET}}$ and leave into drain (see schematic on the right). The Coulomb resonance of the detector appears as a horizontal blue stripe at $V_{\rm{DET}}\sim-803\,\rm{mV}$. b) Cross-section along the dashed line ($V_{\rm{DET}}=-803.8\,\rm{mV}$). Elastic transfer appears as a distinct resonance of negative current, whereas inelastic transfer processes give rise to a shoulder at more negative $V_{\rm{EM}}$. c) Same experiment repeated with negative source potential. Elastic transfer occurs along a red line towards more positive $V_{\rm{EM}}$ and $V_{\rm{DET}}$. Schematic on the right: holes are transferred from the source via the emitter to the detector. Elastic hole transfer gives rise to a resonance of positive current, which clearly appears in the cross-section shown in panel d.}
\end{figure}

Two samples of equivalent geometry were fabricated via standard lithography of a $\rm GaAs/Al_{0.3}Ga_{0.7}As$ heterostructure containing a 2DES with an electron density of $n_{\rm{S}}=2.2\times10^{11}\,\rm{cm^{-2}}$ and a mobility of $\mu=3.4\times10^6\,\rm{cm^{2}V^{-1}s^{-1}}$, residing at the heterointerface $z=90\,\rm{nm}$ underneath the surface. Experimental results were obtained from both samples. Data from sample 1 are shown in figures 1 and 4, data from sample 2 in figures 2 and 3. Exploiting the high symmetry of the gate-induced potential, the direction of electron transfer was swapped between the experiments shown in Figs.~\ref{fig:sample}---\ref{fig:finite-bias} (transfer from left to right) and those shown in Fig.~\ref{fig:hole-transfer} (transfer from right to left). The experiments were carried out using room-temperature electronics connected to the sample mounted in a dilution refrigerator with a base temperature of $T_{\rm{MC}}=9\,\rm{mK}$, resulting in an electronic temperature of the 2DES of typically $T_{\rm{2DEG}}\sim20\,\rm{mK}$.
Applying negative voltages to the Schottky gates defines emitter and detector QDs (see Fig.~\ref{fig:sample}a).

Transport of hot electrons proceeds in a sequence of non-equilibrium tunneling events (Fig.~\ref{fig:sample}b). Electrons are injected from the source via the emitter QD into the reservoir, where they can interact with phonons and cold electrons. A fraction of non-equilibrium electrons reaches the detector QD, where they are energy-selectively detected. In order to characterize the detector ground state (GS), we measure the Coulomb-blockade oscillation of $I_{\rm{DET}}$ as a function of $V_{\rm{DET}}$ while the emitter is in the Coulomb-blockade regime (see Fig.~\ref{fig:sub-thermal}a, schematic I).
To emphasize the thermal broadening, the temperature of the cryostat mixing chamber is set to $T_{\rm{MC}}=83\,\rm{mK}$ (which determines also the phonon temperature of the sample) and the drain bias to $V_{\rm{Drain}}=+2\,\rm{\mu V}$. The tunneling rate $\it{\Gamma}$ and the orbital single-level spacing $\Delta \epsilon$ are tuned such that ${\it{\Delta \epsilon}} > k_{\rm{B}}T_{\rm{2DES}}>({\rm{e}}V_{\rm{Drain}},{h\it{\Gamma}})$. The Coulomb peak is then well described by a purely thermally broadened resonance~\cite{beenakker_theory_1991}:
\begin{equation}
I_{\rm{DET}}=A\times {\rm{cosh}^{-2}}({\rm{e}}\alpha (V_{\rm{DET}}-V_{\rm{OFFSET}})/2k_{\rm{B}}T)
\end{equation}
After determining the lever arm between $eV_{\rm{DET}}$ and energy via finite-bias spectroscopy~\cite{weis_transport_1992} ($\alpha=0.0694$; data not shown), the fit yields the peak amplitude $A$ and a temperature of $T_{\rm{2DES}}=94\,\rm{mK}$. This is in reasonable agreement with the cryostat temperature, considering that sample charge noise and gate noise add to the purely thermal broadening of the Coulomb resonance.

Once the emitter is switched on (by setting $\mu_{\rm{Source}}$ to $200\,\rm{\mu eV}$ and $\mu_{\rm{EM}}$ to $130\,\rm{\mu eV}$; schematic II), a second peak in addition to the Coulomb resonance appears in the detector current at more negative gate voltage and hence at higher energy (Fig.~\ref{fig:sub-thermal}b). This peak arises from the resonance condition $\mu_{\rm{EM}}\approx\mu_{\rm{DET}}$, where electrons are transferred elastically between the QDs. In analogy to transport experiments using tunnel-coupled double QDs~\cite{van_der_vaart_resonant_1995,van_der_wiel_electron_2002}, we expect that the thermal distributions of source, reservoir and drain are irrelevant for the elastically transferred electrons, as $\mu_{\rm{S}}-\mu_{\rm{EM}}\gg k_{\rm{B}}T_{\rm{2DES}}$ and $\mu_{\rm{EM}}\gg k_{\rm{B}}T_{\rm{2DES}}$ (see schematic II). Quantum dots are zero-dimensional objects and have no temperature associated with them, and the acoustic-phonon wavelength at $T\ll 1\,\rm{K}$ is much larger than the QD size. Therefore, the electron--phonon coupling is expected to be reduced compared to free electrons in a 2DES. The line width of the elastically transferred current should be determined by the lifetime broadening of the emitter and detector states, whereas electron--electron and electron--phonon interactions in the reservoir create a background of inelastic processes. We find that a Lorentzian with a full width at half maximum (FWHM) of ${\it{\Delta}}_{\rm{E}}=10.1\,{\rm{\mu eV}}$ provides a reasonable fit to the elastic resonance. In the gate-voltage range between the elastic peak and the Coulomb resonance, the detector probes inelastically scattered electrons (schematic III), which gives access to the non-equilibrium energy distribution in the reservoir arising from partial relaxation of injected electrons. We compare ${\it{\Delta}}_{\rm{C}}$ to ${\it{\Delta}}_{\rm{E}}$ by normalizing both by $3.5k_{\rm{B}}$~\cite{beenakker_theory_1991} for several mixing-chamber temperatures, see Fig.~\ref{fig:sub-thermal}c (the FWHM of a thermally broadened resonance, eq. 1, is $\Delta_{\rm{C}}=3.5\,k_{\rm{B}}T_{\rm{2DES}}$). Whereas ${\it{\Delta}}_{\rm{C}}/3.5k_{\rm{B}}$ changes proportionally  with $T_{\rm{MC}}$ , ${\it{\Delta}}_{\rm{E}}/3.5k_{\rm{B}}$ depends only weakly on temperature within the temperature range accessible in our experiment ($T_{\rm{MC}}=9$--$180\,\rm{mK}$).

For further characterization of our non-equilibrium detection scheme, we mapped the transport properties at increased tunnel-coupling strengths of the emitter and the detector, such that ${h\it{\Gamma}}_{\rm{EM}}\approx {h\it{\Gamma}}_{\rm{DET}}\approx 20\,{\rm{\mu eV}}\gg k_{\rm{B}}T_{\rm{2DES}}$. Figure~\ref{fig:finite-bias}a shows $I_{\rm{EM}}$ as a function of $V_{\rm{EM}}$ and $\mu_{\rm{Source}}$. Distinct parameter regions appear for the Coulomb-blockade regime (white area), for sequential tunneling through the emitter's GS with $I_{\rm{EM}}\approx 1\,\rm{nA}$ (faint-red area), and for the regime of tunneling through an additional excited state (ES) with $I_{\rm{EM}}\approx 2.5\,\rm{nA}$ (dark-red area). By convention, electrons tunneling into the reservoir are defined as positive current. The simultaneously measured $I_{\rm{DET}}$ is shown in Fig.~\ref{fig:finite-bias}b. The detector current $I_{\rm{DET}}$ is around 50 times smaller than $I_{\rm{EM}}$ and negative, meaning that electrons tunnel from the reservoir to drain. Finite $I_{\rm{DET}}$ is only observed within the sequential-tunneling regions of the emitter and it is proportional to the emitter current, as seen in particular by the increase within the ES-emission region (dark blue region in Fig.~\ref{fig:finite-bias}b). A few exemplary emitter--detector configurations are sketched in Fig.~\ref{fig:finite-bias}; resonances between the GSs of emitter and detector (I), between emitter GS and detector ES (II) and between emitter ES and detector GS (III) appear as streaks or steps on top of a background of inelastic current.
Additonal mechanistic insight comes from a comparison of our experimental data to an empirical model. In the regime of strong tunnel coupling and assuming symmetric tunnel coupling to source and drain (${\it{\Gamma}}_{\rm{S}}={\it{\Gamma}}_{\rm{D}}={\it{\Gamma}}/2$), Coulomb resonances are characterized by a lifetime-broadened (and hence Lorentzian) spectral transmission probability~\cite{foxman_effects_1993,datta_electronic_1997}
\begin{equation}
T_{\rm{EM,DET}}(E)=\frac{({\it{h\Gamma}})^2/2}{({\it{h\Gamma}}/2)^2+(E-E_{\rm{0}})^2}
\end{equation}
where the tunneling rates and energies have been independently determined by fitting eq. (2) to the Coulomb resonances of emitter and detector in the linear response regime (data not shown). The tunnel coupling of the emitter and detector ground state is ${\it{h\Gamma}}_{\rm{EM}}^{\rm{GS}}\approx{\it{h\Gamma}}_{\rm{DET}}^{\rm{GS}}\approx20\,{\rm{\mu eV}}\gg k_{\rm{B}}T\approx2\,\rm{\mu eV}$ for $T_{\rm{2DES}}=25\,\rm{mK}$, which means that temperature broadening is negligible. The tunnel coupling of the emitter excited state is ${\it{h\Gamma}}_{\rm{EM}}^{\rm{ES}}\approx40\,{\rm{\mu eV}}$ and gives rise to the strong increase of the emitter current seen in Fig.~\ref{fig:finite-bias}a) as soon as the excited state enters the bias windows. The resonance energies $E_0$ are given either by the emitter GS ($\mu_{\rm{EM}}$), the emitter ES ($\mu_{\rm{EM}}^*$), or the detector GS ($\mu_{\rm{DET}}$), respectively. For the sake of simplicity, we did not include detector ESs in our model. In addition, double occupation of a QD is neglected as the charging energy $U\approx1\,\rm{meV}\gg {\it{h\Gamma}}$. The spectrally resolved transmission through the emitter was calculated via a rate-equation approach~\cite{beenakker_theory_1991} which, after an extension to a two-level energy scheme, yields $T_{\rm{EM}}(E)=\overline{p}^{\rm{ES}}T_{\rm{EM}}^{\rm{GS}}(E)+\overline{p}^{\rm{GS}}T_{\rm{EM}}^{\rm{ES}}(E)$ where $\overline{p}^{\rm{GS}}$ denotes the probability that the emitter GS is not occupied and $\overline{p}^{\rm{ES}}$ is the probability that the emitter ES is not occupied. If interference effects in the region between emitter and detector and relaxation processes in the QDs are neglected, the elastically transmitted current can be expressed through the convolution of the spectral current density of the emitter GS and ES with that of the detector GS  within the accessible bias window: $I_{\rm{ELAST}}\propto \frac{e}{h}\int\limits_{0}^{\mu_{\rm{S}}}T_{\rm{EM}}(E) \, T_{\rm{DET}}(E) \, dE$. We assumed that inelastic processes are dominated by electron--electron interaction, as the excess energies $E\lesssim 1\,\rm{meV}$~\cite{giuliani_lifetime_1982} are relatively small. Therefore we expect that an electron injected at energy $E$ relaxes to another energy $\epsilon$ during the transfer to the detector with probability $P(\epsilon|E)\,d\epsilon\propto(E-\epsilon)^2\,d\epsilon$. The inelastic contribution to the detected current is then given by $I_{\rm{INELAST}}\propto \frac{e}{h}\int\limits_{0}^{\mu_{\rm{S}}}\int\limits_{0}^{\epsilon}T_{\rm{EM}}(E) \, P(\epsilon|E) \, T_{\rm{DET}}(E) \, d\epsilon \, dE$.
We find good agreement between our experimental data and the calculated total current $I_{\rm{\Sigma}}=I_{\rm{ELAST}}+I_{\rm{INELAST}}$ when choosing a ratio of $I_{\rm{ELAST}}/I_{\rm{INELAST}}=20$. Figure~\ref{fig:finite-bias}c) shows the calculated current $I_{\rm{\Sigma}}$ as a function of ${\it{\Delta}} V_{\rm{EM}}$ and $\mu_{\rm{Source}}$ with the detector GS at $\mu_{\rm{DET}}=0.1\,\rm{meV}$ (left) and $\mu_{\rm{DET}}=0.3\,\rm{meV}$ (right). As the detector energy is increased, the onset of finite current moves to larger $\mu_{\rm{Source}}$. The large triangular region determined by current through the emitter GS and the smaller triangle (top left) given by the emitter ES are reflected in the dependence of the calculated current on $V_{\rm{EM}}$ and $\mu_{\rm{Source}}$. Both in the experimental data and in the model, elastically transferred electrons and inelastically scattered electrons can be clearly discriminated (cf. supplementary material).

Finally, using QDs as energy-selective emitters and detectors offers another distinct advantage over QPC-based spectrometry: QPCs suppress hole transmission because the potential barrier is much wider underneath the Fermi energy (for hot holes) than above the Fermi energy (for hot electrons). This limitation does not exist for transport governed by Coulomb blockade (see Fig.~\ref{fig:hole-transfer}a).

The region of elastic electron transfer is clearly visible when $V_{\rm{EM}}$ and $V_{\rm{DET}}$ are varied simultaneously. A cross-section at constant $V_{\rm{DET}}$ (Fig.~\ref{fig:hole-transfer}b) reveals again elastic and inelastic transfer of non-equilibrium electrons. When the source potential is negative --- that is, below the Fermi energy of the reservoir and the drain --- the resonance condition between emitter and detector moves to more positive $V_{\rm{EM}}$ and $V_{\rm{DET}}$, as shown in Fig.~\ref{fig:hole-transfer}c. Elastic charge transfer now generates a positive current, indicating elastic transfer of non-equilibrium holes. A cross-section at constant $V_{\rm{DET}}$ (Fig.~\ref{fig:hole-transfer}d) shows that also for holes, elastic and inelastic transfer events can be clearly discriminated.

In conclusion, we have investigated equilibrium and non-equilibrium electron transfer via two spatially separated QDs. Resonant and non-resonant transfer processes can be observed with a spectral resolution of few $\rm{\mu eV}$. We find that the spectral width of the resonant elastic transfer peak can be well described by the life-time broadening of the two QDs and exhibits only a weak dependence on temperature. A model involving tunnel-broadened resonances of the emitter and detector reproduces most experimental features, but additional theoretical work is required to be able to quantify the relaxation rates of hot electrons. Furthermore, we have demonstrated the ability of our spectrometry scheme to emit and detect both non-equilibrium electrons and non-equilibrium holes. For future studies, transfer experiments could be repeated at much lower tunneling rates, ${\it{\Gamma}}\lesssim 10\,\rm{kHz}$, with adjacent charge detectors, in order to gather statistics of individual electron-transfer events~\cite{gustavsson_counting_2006}. Measuring the charge of a single electron in two QDs consecutively might offer access to the implementation of a conditional measurement protocol in a solid-state device~\cite{aharonov_how_1988}. One might also envision that co-tunneling processes can be measured and characterized in a time-resolved manner~\cite{romito_weak_2013}.

We acknowledge the support of the ETH FIRST laboratory and financial support of the Swiss Science Foundation (Schweizerischer Nationalfonds, NCCR Quantum Science and Technology).

\bibliography{Bibliothek}

\end{document}



\section{Supplementary: Comparison between experimental data and model}
\begin{figure}[b]
\includegraphics[scale=1]{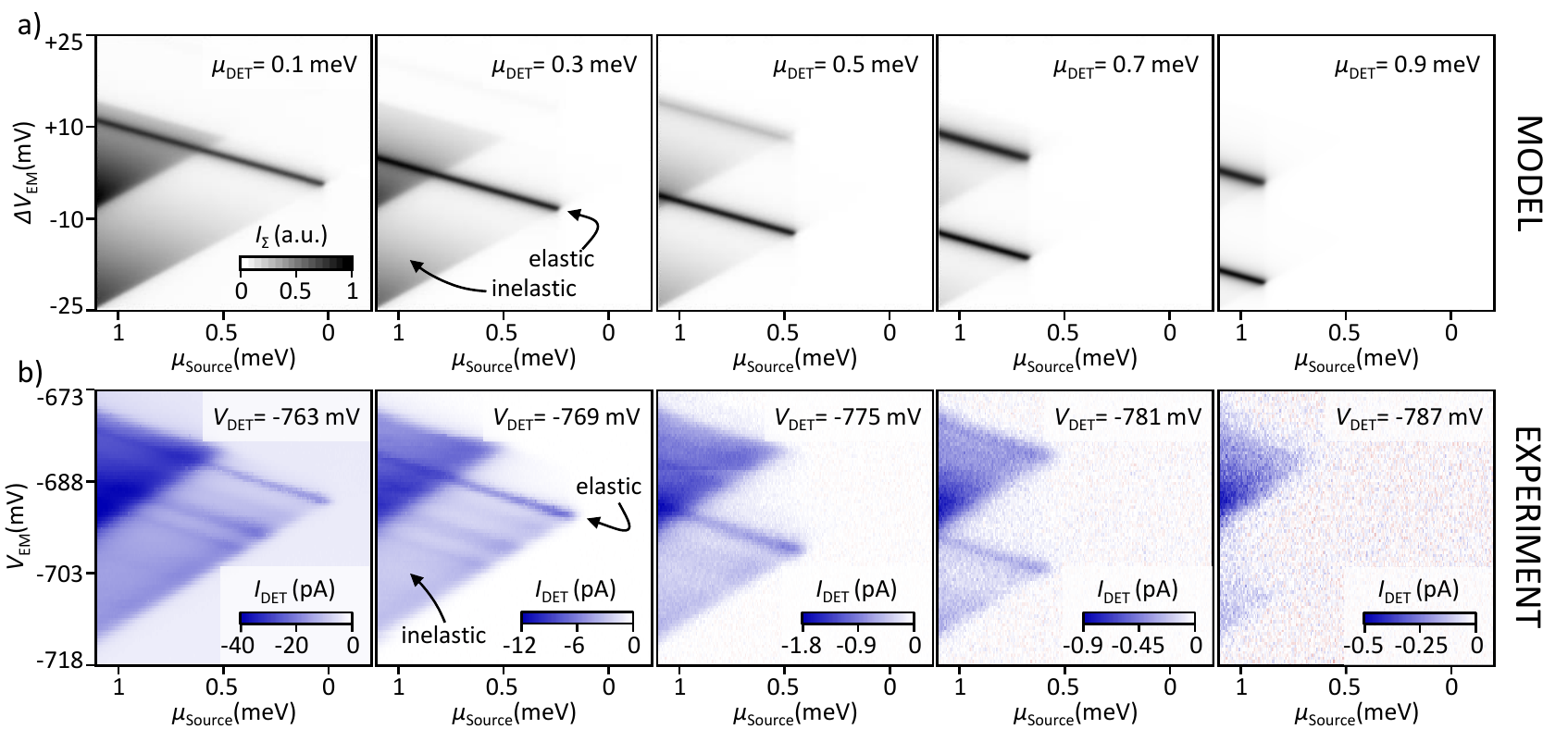}
\caption{\label{fig:model-experiment} a) Calculated current through the two-QD system, plotted in linear gray scale as a function of ${\it{\Delta}} V_{\rm{EM}}$ and $\mu_{\rm{Source}}$. Resonance between emitter GS and detector GS occurs along the black streak, current within the filled triangles is caused by inelastic scattering events in the reservoir. The individual panels show setting with different energies of the detector GS (increasing from left to right). At higher detector-GS energies, the onset of transmission shifts to larger $\mu_{\rm{Source}}$. b) Measured $I_{\rm{DET}}$ in linear color scale, plotted as a function of $V_{\rm{EM}}$ and $\mu_{\rm{Source}}$. The individual panels again show setting with different energies of the detector GS, increasing from left to right in steps of ${\it{\Delta}} \mu_{\rm{DET}}=\alpha {\it{\Delta}} V_{\rm{DET}}\approx200\,\rm{\mu eV}$. As the detector GS is increased, fewer elastic and inelastic electrons are detected.}
\end{figure}

\bibliography{Bibliothek}